\documentclass[10pt,letterpaper]{article}
\usepackage{opex3}
\usepackage{color}
\usepackage[colorlinks=true,urlcolor=blue,linkcolor=black,citecolor=black,breaklinks=true,bookmarks=false]{hyperref}
\usepackage{amsmath}
\usepackage{amssymb}
\usepackage{graphicx}
\usepackage{braket}
\usepackage{mathtools}

\hypersetup{pdfauthor={J. Oppermann, J. Straubel, I. Fernandez-Corbaton, C. Rockstuhl},
pdftitle={A normalization approach for scattering modes to be of use in classical and quantum electrodynamics}}

\newcommand{\ii}{\text{i}}
\newcommand{\ee}{\text{e}}
\renewcommand{\vec}[1]{\mathbf{#1}}
\let\oldhat\hat
\renewcommand{\hat}[1]{\oldhat{\vec{#1}}}

\begin{document}

\title{A normalization approach for scattering modes to be of use in classical and quantum electrodynamics}

% NOT FOR RELEASE
%\author{\textcolor{Jens}{J.~Oppermann}\textsuperscript{1},
%\textcolor{Jakob}{J.~Straubel}\textsuperscript{1},
%\textcolor{Ivan}{I.~Fernandez-Corbaton}\textsuperscript{2},
%\textcolor{Carsten}{C.~Rockstuhl}\textsuperscript{1,2}}
% END

% RELEASE VERSION
\author{J.~Oppermann\textsuperscript{1}, J.~Straubel\textsuperscript{1}, I.~Fernandez-Corbaton\textsuperscript{2}, C.~Rockstuhl\textsuperscript{1,2}}
% END

\address{\textsuperscript{1}Institute of Theoretical Solid State Physics, Karlsruhe Institute of Technology, Wolfgang-Gaede-Str.~1, D-76131 Karlsruhe, Germany}

\address{\textsuperscript{2}Institute of Nanotechnology, Karlsruhe Institute of Technology, Hermann-von-Helmholtz-Platz~1, D-76021 Eggenstein-Leopoldshafen, Germany}

\email{jens.oppermann@kit.edu} %% email address is required



\begin{abstract*}
We propose a novel scheme to normalize scattering modes of the electromagnetic field. By relying on analytical solutions for Maxwell's equations in the homogenous medium outside the scatterer, we derive normalization conditions that only depend on the electromagnetic field on the surface of a sphere containing the scatterer. We pay special attention to the important cases of plane wave illumination and illumination with a multipolar field, for which an explicit and easy to use normalization condition is derived. We demonstrate the versatility of our method by normalizing scattering modes of some selected metallic and dielectric scatterers of different geometries in the context of different application scenarios. Since every quantum mechanical treatment of light-matter interaction requires the proper normalization of electromagnetic fields, we deem our proposed normalization scheme broadly applicable independent of the scatterer involved.
\vspace{1mm}
\end{abstract*}

\ocis{(030.4070) Modes; (270.0270) Quantum optics; (270.5580) Quantum electrodynamics; (290.5825) Scattering theory.} % REPLACE WITH CORRECT OCIS CODES FOR YOUR ARTICLE, MINIMUM OF TWO; Avoid using the OCIS codes for “General” or “General science” whenever possible.
%For a complete list of OCIS codes, visit: https://www.osapublishing.org/oe/submit/ocis/

%%%%%%%%%%%%%%%%%%%%%%% References %%%%%%%%%%%%%%%%%%%%%%%%%

\bibliography{references}

%%%%%%%%%%%%%%%%%%%%%%%%%%  body  %%%%%%%%%%%%%%%%%%%%%%%%%%

\section{Introduction}

Cavity quantum electrodynamics explores the interaction between light, which possesses potentially non-classical properties, and matter in the presence of dielectric or plasmonic cavities. However, in a slightly broader sense, any spatial distribution of classical matter in a finite region of space affecting the propagation of electromagnetic fields can be considered a cavity. The cavity might simply have a very poor quality factor. The field of cavity quantum electrodynamics combines classical and quantum mechanical aspects, in that the relevant modes of the electromagnetic field are determined classically and the coupling to matter is described quantum mechanically. This approach has led to many exciting applications, such as single photon sources \cite{McKeever2003, Press2007, Schietinger2009, Claudon2010, Esteban2010, Birowosuto2012, Busson2012, Filter2014, Nowak2014, Straubel2016} and the generation of squeezed light \cite{Michler2000, Lounis2000, Safavi2013, Martin-Cano2014}. The process of linking classical and quantum calculations, however, can be rather challenging. 

One of the central problems is the correct normalization of electromagnetic field modes. Since this process involves integration over an infinitely extended spatial domain \cite{Vogel1994}, many common simulation techniques such as the finite element method (FEM) or the finite-difference time-domain (FDTD) method can't cope with it. Furthermore, even in cases with known analytical solutions and feasible integrations, the correct normalization is usually not obvious since multiple $\delta$-distributions can arise. As a consequence, the normalization of modes is usually discussed in terms of eigenmodes. Eigenmodes are self-consistent solutions to Maxwell's equations for a given material distribution in space in the absence of a source~ \cite{jackson1975electrodynamics, desoer2009basic, Huttner1991}. In every basic course on quantum optics, monochromatic, elliptically polarized plane waves are discussed as the eigenmodes of the homogenous isotropic space \cite{grynberg2010introduction, mandel1995optical}. Their normalization starts with the consideration of some finite or periodic quantization volume, that allows easily for a proper normalization \cite{loudon2000quantum, haroche2006exploring}. Such normalization of the field requires to set a field amplitude such that the energy density contained in the mode integrated across the considered volume corresponds to that of a single photon with an energy associated to the frequency of the considered mode \cite{walls2007quantum, Fetter}. In a final step, this quantization volume is extended to infinity \cite{scully1999quantum}. Such quantization procedure can be also applied to slightly more complicated optical systems such as dielectric waveguides \cite{teich1991fundamentals} or photonic crystals \cite{joannopoulos2011photonic, de2014roundtrip}, where the eigenmodes are invariant or at least Bloch-periodic in at least one direction. Moreover, eigenmodes of isolated scatterers \cite{bryant2008mapping, bai2013efficient, kristensen2013modes, makitalo2014modes, doost2014resonant, ge2015quasi, vial2016coupling} and their quantization \cite{ge2015quantum} are currently intensively explored.

\begin{figure}
\begin{center}
\includegraphics[width=0.85\textwidth]{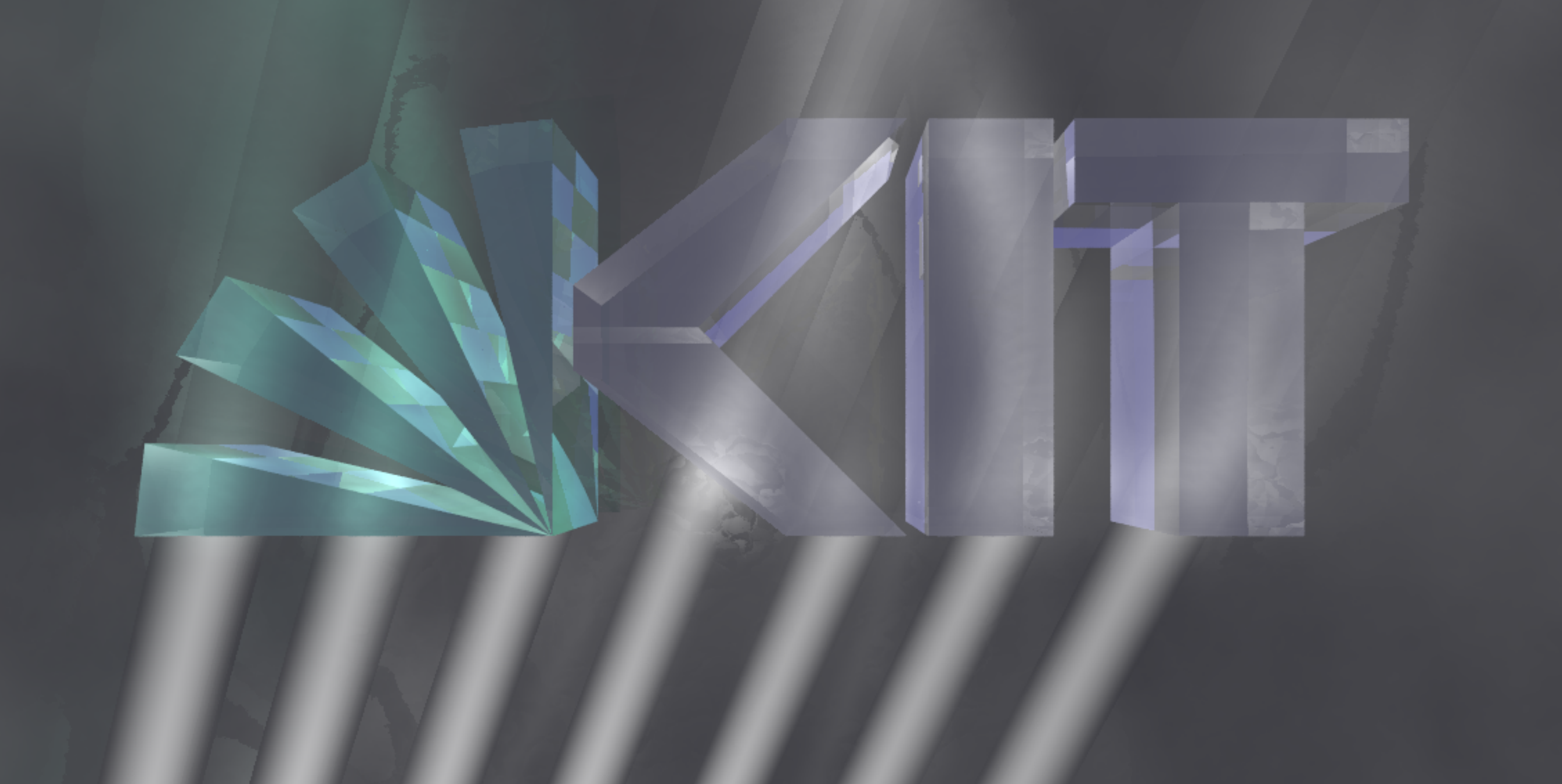}
\end{center}
\caption{Localized scatterers of arbitrary shape can be described with the proposed formalism. \label{fig:kit_scatteringPIC}}
\end{figure}

However, such quantization does not resemble many practical problems of interest. There, a given source of nonclassical light is usually considered, that shall interact with some material distribution in space. A pivotal example that is an essential part in any textbook on quantum optics would be the operation of a beam splitter \cite{schleich2011quantum}. How does one have to normalize the field of a single photon interacting with such a beam splitter? A slightly more illustrative example is shown in Fig. 1, where an arbitrarily looking scatterer is illuminated by an external field. In the spirit of a second quantization approach, it is usually suggested that we need to know at first the normalized electromagnetic modes of the homogenous (or empty) space containing additionally the beam splitter or any other scatterer in a manner exactly analogous to obtaining plane waves as the modes of the free space. However, how these modes shall be obtained and, in particular, normalized has not yet been described.

Indeed, besides the free space only incrementally more complicated scenarios have been considered when it comes to the quantization and particularly the normalization of the fields caused by a given illumination in the presence of some matter. Here, the pioneering work of Carniglia and Mandel needs to be mentioned. They quantized and normalized the field generated by a plane wave that is reflected and transmitted at the interface that divides the space into two semi-infinite half spaces \cite{Carniglia1971}. They did so by considering the triplet of the incident, the reflected, and the scattered plane wave, i.e. the entire solution to the scattering problem of that very specific situation, in the normalization simultaneously. This approach was extended by others \cite{Khosravi1991, Rigneault1997, vial2014quasimodal} but eventually it had been limited to the scenario of a plane wave illuminating some sort of stratified media. No other situation has been considered until now. This is surprising considering the importance of the basic scenario in quantum optics, where a nonclassical state of light interacts with some material distribution that is finite in space.

The purpose of the present contribution is to close this gap. We provide here a novel normalization approach for scattering modes that arise if a given illumination interacts with a localized object. This normalization approach is of use in both classical and quantum electrodynamics. The key to success will be the derivation of a formulation that only requires surface integrations and not an integration across the infinite space to arrive at a proper normalization condition. The scheme is applicable to arbitrary scattering geometries as long as the scatterer is of finite size. In Section \ref{sec:scheme} we develop the normalization scheme and consider the cases of multipole and plane wave illumination. We provide easy-to-implement formulas requiring at most an integration over a solid angle. In Section \ref{sec:numerics} we discuss different numerical examples and demonstrate the calculation of several important quantum optical quantities. We summarize our findings in Section \ref{sec:conclusion}.

\section{Normalization scheme}
\label{sec:scheme}

\begin{figure}
\begin{center}
\includegraphics[width=.4\textwidth]{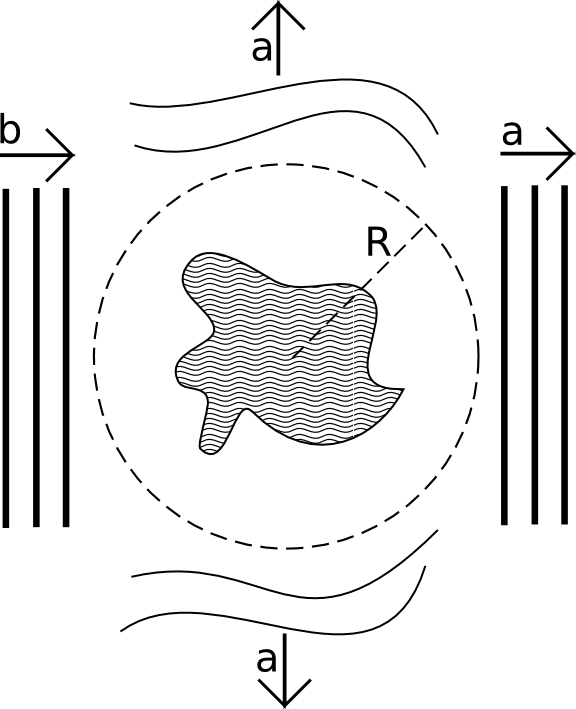}
\end{center}
\caption{Sketch of the system under consideration. An electromagnetic wave is incident on a localized scatterer, giving rise to a scattered field. The cavity can be completely encased in a virtual sphere of radius $R$, outside of which the field propagates freely. The arrows indicate the flow of energy with a and b marking outgoing and incoming contributions, respectively.}
\label{fig:scatter_geometry}
\end{figure}

We consider the system schematically shown in Fig.~\ref{fig:scatter_geometry}. A localized scatterer is embedded in a lossless background material and illuminated by a propagating electromagnetic radiation field. This gives rise to an outgoing scattered field. Since the scatterer shall be localized in space, we can always choose a sphere around the center of coordinates that completely containes the scatterer. We stress the fact that we do not impose an upper bound on the size of the sphere, but it must be finite. This sphere divides the space into two regions, one inhomogeneous and the other one homogeneous.

Our goal is to find a complete set of scattering solutions $\vec{E}_{ka}(\vec{r})$ at a given frequency that are properly orthonormalized to contain the energy of a single photon according to \cite{Landau1984, Vogel1994}
\begin{eqnarray}\label{eq:normalisation_integral}
\int d^{3}rS_{k}(\vec{r})\vec{E}_{ka}^{\dagger}(\vec{r})\vec{E}_{k'a'}(\vec{r}) &=& \frac{\hbar\omega}{2\epsilon_{0}}\delta(k-k')\delta_{aa'} \quad , \\
S_k(\vec{r}) &=& \frac{\text{d}}{\text{d}\omega}\left[\omega\text{Re}\left(\epsilon(\vec{r},\omega)\right)\right] \quad .
\end{eqnarray}
Here, the scattering modes are labeled according to their wavenumber $k$ and their angular distribution, described by a generic index $a$. Please note that $\vec{E}_{ka}(\vec{r})$ denotes the total electromagnetic field (as opposed to the scattered field).

The scheme proposed here relies on an expansion of the far field in terms of solutions of Maxwell's equations in homogeneous media. The most widely used systems of solutions are vector spherical harmonics and plane waves and we will therefore limit our discussion to those two expansions.

\subsection{Expansion in spherical harmonics}
\label{sec:spherical_harmonics}

It is well known that the electromagnetic wave equation for monochromatic waves in a homogenous medium
\begin{equation}
\vec{\nabla}\times\vec{\nabla}\times\vec{E}\left(\vec{r}\right) - k^2(\omega)\vec{E}\left(\vec{r}\right) = 0 \quad , \label{eq:wave_equation}
\end{equation}
where $k^2(\omega) = \epsilon_0\epsilon_b(\omega)\omega^2/c^2$, has the general solution \cite{Bohren1998, Kristensson2016}
\begin{align}
\vec{E}\left(\vec{r}\right)  = \sum_{l=1}^\infty\sum_{m=-l}^l \Biggl[ &
a_{1lm}h_l^{(1)}\left(kr\right)\vec{A}_{1lm}\left(\hat{r}\right) \nonumber\\ & +
a_{2lm}\left[\left(g_l^{(1)}\left(kr\right) + \frac{h_l^{(1)}\left(kr\right)}{kr}\right)\vec{A}_{2lm}\left(\hat{r}\right) + l(l+1) \frac{h_l^{(1)}\left(kr\right)}{kr}\vec{A}_{3lm}\left(\hat{r}\right)\right] \nonumber\\ & +
b_{1lm}h_l^{(2)}\left(kr\right)\vec{A}_{1lm}\left(\hat{r}\right) \nonumber\\ & +
b_{2lm}\left[\left(g_l^{(2)}\left(kr\right) + \frac{h_l^{(2)}\left(kr\right)}{kr}\right)\vec{A}_{2lm}\left(\hat{r}\right) + l(l+1) \frac{h_l^{(2)}\left(kr\right)}{kr}\vec{A}_{3lm}\left(\hat{r}\right)\right] \Biggr] \quad , \label{eq:spherical_harmonics_solution}
\end{align}
where $h_l^{(n)}$ and $g_l^{(n)}$ denote the spherical Hankel functions of the $n$-th kind and their first derivatives, respectively, and $\vec{A}_{nlm}$ are the vector spherical harmonics
\begin{eqnarray}
\vec{A}_{1lm}\left(\hat{r}\right) & = & \frac{1}{\sqrt{l(l+1)}}\vec{\nabla}Y_{lm}\left(\hat{r}\right)\times\vec{r} \quad , \nonumber \label{eq:spherical_harmonics1} \\
\vec{A}_{2lm}\left(\hat{r}\right) & = & \frac{1}{\sqrt{l(l+1)}}r\vec{\nabla}Y_{lm}\left(\hat{r}\right) \quad , \nonumber \label{eq:spherical_harmonics2} \\
\vec{A}_{3lm}\left(\hat{r}\right) & = & \frac{1}{\sqrt{l(l+1)}}\hat{r}Y_{lm}\left(\hat{r}\right) \quad , \label{eq:spherical_harmonics3}
\end{eqnarray}
where $Y_{lm}$ are the (scalar) spherical harmonics and $\hat{r}$ denotes the normalized position vector. Please note that the coefficients labeled $a_{nlm}$ and $b_{nlm}$ correspond to outgoing and incoming waves, respectively, and that the index $n$ determines the field polarization. If the dielectric function $\epsilon(\vec{r},\omega)$ varies in space, the above solution no longer holds. In the case of a localized scatterer, however, a virtual sphere of radius $R$ that contains the scatterer can always be identified that is embedded in a homogenous environment of dielectric function $\epsilon_b(\omega)$. The above solution then applies outside of this sphere.

A scattering problem is hence defined by the wavenumber $k$, the scattering geometry, and the incoming electromagnetic far field described by the parameters $b_{nlm}$ in Eq.~\eqref{eq:spherical_harmonics_solution}. The solution to such a scattering problem is a set of values for the parameters $a_{nlm}$, which describe the outgoing electromagnetic field. Please note that the outgoing field consists of a scattered part and a part similar to the incoming field that is transmitted without interaction. The total field is given by the incoming plus the outgoing field.

Assume now that we know the solution of the scattering problem in a finite domain encasing the scatterer, e.g.~from finite element calculations. By using the orthogonality relations
\begin{equation}
\int_{\text{Sphere}} d\Omega \vec{A}_{nlm}^\dagger\left(\hat{r}\right)\cdot\vec{A}_{n'l'm'}\left(\hat{r}\right) = C_{nl}\delta_{nn'}\delta_{ll'}\delta_{mm'}\label{eq:spherical_harmonics_orthogonality}
\end{equation}
with
\begin{eqnarray}
C_{1l} &=& C_{2l} = 1 \quad \mathrm{and} \nonumber \\
C_{3l} &=& \frac{1}{l(l+1)}
\end{eqnarray}
we can easily reach the following expressions for the coefficients $a_{nlm}$  in Eq.~\eqref{eq:spherical_harmonics_solution}: 
\begin{eqnarray}
a_{1lm} &=& \frac{1}{h_l^{(1)}\left(kr\right)}\int d\Omega \vec{A}_{1lm}^\dagger\left(\hat{r}\right) \cdot \vec{E}\left(\vec{r}\right) - b_{1lm}\frac{h_l^{(2)}\left(kr\right)}{h_l^{(1)}\left(kr\right)} \quad ,\nonumber \\
a_{2lm} &=& \frac{1}{g_l^{(1)}\left(kr\right)+h_l^{(1)}\left(kr\right)/kr}\int d\Omega \vec{A}_{2lm}^\dagger\left(\hat{r}\right) \cdot \vec{E}\left(\vec{r}\right) - b_{2lm}\frac{g_l^{(2)}\left(kr\right)+h_l^{(2)}\left(kr\right)/kr}{g_l^{(1)}\left(kr\right)+h_l^{(1)}\left(kr\right)/kr} \quad .
\end{eqnarray}
The last terms serve to subtract the incoming field from the total field, leaving only the outgoing field. We proceed by choosing a complete (but not necessarily orthogonal) system of scattering solutions. To keep matters simple, we choose modes $\vec{E}_{knlm}(\vec{r})$ with wavenumber $k$ for which only one parameter $b_{nlm}$ is different from zero. This means that the scatterer has been illuminated by a single multipole field, and $\vec{E}_{knlm}$ is the total field in this scattering problem. The integrals to be evaluated read [following Eq.~\eqref{eq:normalisation_integral}]
\begin{equation} \label{eq:norm_integral}
W^{(k,n,l,m,k',n',l',m')} = \int d^{3}rS_k(\vec{r})\vec{E}_{knlm}^{\dagger}(\vec{r})\cdot\vec{E}_{k'n'l'm'}(\vec{r}) \quad .
\end{equation}
We now truncate the exact solution at a value $l = l_{max}$ of the angular momentum index and subsequently choose a virtual sphere of radius $R$ in such a way that $kR >> l_{max}^2$ is satisfied. At the end of our calculation we will take the limit $R\rightarrow\infty$, hence restoring all orders in $l$. This allows us to use the asymptotic forms of the Hankel functions and their derivatives in the outer region where $r>R$ \cite{Bohren1998}
\begin{eqnarray}
h_l^{(1)}(kr) &\xrightarrow{r\rightarrow\infty}& (-\ii)^{l+1} \frac{\ee^{\ii kr}}{kr} \quad , \nonumber \\
g_l^{(1)}(kr) &\xrightarrow{r\rightarrow\infty}& (-\ii)^{l} \frac{\ee^{\ii kr}}{kr} \quad , \nonumber \\
h_l^{(2)}(kr) &\xrightarrow{r\rightarrow\infty}& (\ii)^{l+1} \frac{\ee^{-\ii kr}}{kr} \quad , \nonumber \\
g_l^{(2)}(kr) &\xrightarrow{r\rightarrow\infty}& (\ii)^{l} \frac{\ee^{-\ii kr}}{kr} \quad . \label{eq:hankel_asymptotic}
\end{eqnarray}
Insertion of these expressions into Eq.~\eqref{eq:spherical_harmonics_solution} yields the asymptotic form of the outside field
\begin{equation}
\vec{E}_{knlm}(r>R) \approx \vec{E}_{knlm}^{(\text{ff})}(\vec{r}):=\vec{A}^{(\text{out})(k,n,l,m)}(\hat{r})\frac{\ee^{\ii kr}}{kr}+\vec{A}_{nlm}^{(\text{in})}(\hat{r})\frac{\ee^{-\ii kr}}{kr} \quad ,
\end{equation}
where the angular components are
\begin{eqnarray}
\vec{A}^{(\text{out})(k,n,l,m)}(\hat{r}) &=& \sum_{l'm'}\left[(-\ii)^{l'+1}a_{1l'm'}^{(k,n,l,m)}\vec{A}_{1l'm'}(\hat{r})+(-\ii)^{l'}a_{2l'm'}^{(k,n,l,m)}\vec{A}_{2l'm'}(\hat{r})\right] \quad , \\
\vec{A}_{nlm}^{(\text{in})}(\hat{r}) &=& \ii^{l+\delta_{n1}}b_{nlm}\vec{A}_{nlm}(\hat{r}) \quad .
\end{eqnarray}
We now proceed by splitting the integration region of \eqref{eq:norm_integral} into outside and inside parts
\begin{equation}
W^{(k,n,l,m,k',n',l',m')} = \overline{W}^{(k,n,l,m,k',n',l',m')} + F^{(k,n,l,m,k',n',l',m')}\left(R\right) \quad . \label{eq:integral_full}
\end{equation}
Here, the approximation \eqref{eq:hankel_asymptotic} is applied outside of the sphere with radius $R$. The terms in Eq.~\eqref{eq:integral_full} then read
\begin{eqnarray}
\overline{W}^{(k,n,l,m,k',n',l',m')} &\approx& \epsilon_b\int d^3r \vec{E}_{knlm}^{(\text{ff})\dagger}(\vec{r})\cdot\vec{E}_{k'n'l'm'}^{(\text{ff})}(\vec{r}) \qquad \mathrm{and} \\
F^{(k,n,l,m,k',n',l',m')}\left(R\right) &=& \int_{r<R}d^3r\bigl[S_k(\vec{r})\vec{E}_{knlm}^{\dagger}(\vec{r})\cdot\vec{E}_{k'n'l'm'}(\vec{r}) - \epsilon_b\vec{E}_{knlm}^{(\text{ff})\dagger}(\vec{r})\cdot\vec{E}_{k'n'l'm'}^{(\text{ff})}(\vec{r})\bigr] \quad .
\end{eqnarray}
Please note that only $F$ depends on the radius $R$ of the virtual sphere. Furthermore, since $F$ is a finite integral over a well-behaved function, $F$ is finite. Later we will see that the finiteness of $F$ leads to it being infinitely smaller than $W$, which means that neither the radius $R$ of the virtual sphere nor the electromagnetic field distribution within it have any influence on the field normalization.

For the evaluation of the integral $\overline{W}$ we use the following identity
\begin{equation}
\int_{0}^{\infty}dre^{iKr}=\pi\delta(K)+\text{P}\frac{i}{K} \quad ,
\end{equation}
where $\text{P}$ denotes the principal part \cite{Bronshtein2013}. This immediately leads us to
\begin{eqnarray}
\frac{kk'}{\epsilon_b}\overline{W}^{(k,n,l,m,k',n',l',m')} &=& Z_{\text{io}}^{(k,n,l,m,n',l',m')}[\pi\delta(k+k')+\text{P}\frac{i}{k+k'}] \nonumber \\
& & +Z_{\text{oi}}^{(n,l,m,k',n',l',m')}[\pi\delta(-k-k')+\text{P}\frac{i}{-k-k'}] \nonumber \\
& & +Z_{\text{ii}}^{(n,l,m,n',l',m')}[\pi\delta(k-k')+\text{P}\frac{i}{k-k'}] \nonumber \\
& & +Z_{\text{oo}}^{(k,n,l,m,k',n',l',m')}[\pi\delta(-k+k')+\text{P}\frac{i}{-k+k'}] \quad ,\label{eq:norm_integral_infinite_part}
\end{eqnarray}
with the terms
\begin{eqnarray}
Z_{\text{oi}}^{(k,n,l,m,n',l',m')} &=& \int d\Omega \vec{A}^{(\text{out})(k,n,l,m)\dagger}(\hat{r}) \cdot \vec{A}_{n'l'm'}^{(\text{in})}(\hat{r})  \quad , \nonumber \\ 
Z_{\text{io}}^{(n,l,m,k',n',l',m')} &=& \int d\Omega \vec{A}_{nlm}^{(\text{in})\dagger}(\hat{r}) \cdot \vec{A}^{(\text{out})(k',n',l',m')}(\hat{r})  \quad , \nonumber \\ 
Z_{\text{ii}}^{(n,l,m,n',l',m')} &=& \int d\Omega \vec{A}_{nlm}^{(\text{in})\dagger}(\hat{r}) \cdot \vec{A}_{n'l'm'}^{(\text{in})}(\hat{r})  \quad , \nonumber \\ 
Z_{\text{oo}}^{(k,n,l,m,k',n',l',m')} &=& \int d\Omega \vec{A}^{(\text{out})(k,n,l,m)\dagger}(\hat{r}) \cdot \vec{A}^{(\text{out})(k',n',l',m')}(\hat{r}) \quad .
\end{eqnarray}
We now turn to the central point of this argumentation. In the above derivations we have only considered solutions corresponding to a definite value of the wavenumber $k$. Such solutions, however, extend across all space and carry infinite energy, making them clearly unphysical. Physical solutions can in general be obtained by linear superposition, e.g.
\begin{equation} \label{eq:physical_field}
\vec{E}_{nlm}(\vec{r}) = \int_0^\infty dk f(k) \vec{E}_{knlm}(\vec{r}) \quad .
\end{equation}
We note here that the function $f(k)$ can be assumed to be real-valued, since any phase factors can be absorbed into the solutions $\vec{E}_{knlm}(\vec{r})$. We will now assume that $f(k)$ only has support on a wavenumber range of width $\Delta k$ centered around $k_0$. At the end of the calculation we will take the limit $\Delta k \rightarrow 0$ to restore the limit of definite wavenumbers.

Inserting the field \eqref{eq:physical_field} into the normalization integral \eqref{eq:normalisation_integral} and approximating $S_{k}(\vec{r})$ by its value at the central wavenumber $k_0$ we obtain
\begin{eqnarray}
\int d^3r S_{k_0}(\vec{r})\vec{E}_{nlm}^\dagger(\vec{r})\cdot\vec{E}_{n'l'm'}(\vec{r}) &\approx& \int_0^\infty dk \int dk' f(k) f(k')  W^{(k,n,l,m,k',n',l',m')} \quad ,\label{eq:physical_field_norm}
\end{eqnarray}
which becomes exact in the limit $\Delta k \rightarrow 0$.

We will now discuss the contributions of the individual terms in Eq.~\eqref{eq:norm_integral_infinite_part} to the integral \eqref{eq:physical_field_norm}. Since $\Delta k << k_0$ can be assumed, the terms involving $\delta(\pm(k+k'))$ vanish and the ones involving $\pm\ii\text{P}(k+k')^{-1}$ remain finite, so we can absorb them in $F$. The thus modified finite contribution is denoted $\overline{F}$. The terms involving $\pm\ii\text{P}(k-k')^{-1}$ on the other hand can grow without limit, but they are antisymmetric under $k \leftrightarrow k'$ while the rest of the integrand in Eq.~\eqref{eq:physical_field_norm} is symmetric. Their contribution therefore vanishes identically. This leaves us with integrals over $\delta$-distributions and a finite contribution, reading
\begin{eqnarray}
\int d^3r S_{k_0}(\vec{r})\vec{E}_{nlm}(\vec{r})\vec{E}_{n'l'm'}(\vec{r}) &\approx& \int_0^\infty dk f^2(k) \frac{\pi\epsilon_b}{k^2} \left(Z_{\text{ii}}^{(n,l,m,n',l',m')}+Z_{\text{oo}}^{(k,n,l,m,k',n',l',m')}\right) + \nonumber \\
& & \int_0^\infty dk \int_0^\infty dk' f(k) f(k') \overline{F}^{(k,n,l,m,k',n',l',m')}(R) \quad . \label{eq:norm_integral_regularized}
\end{eqnarray}
If we now choose a value of $\Delta k$ much smaller than the typical variation length of the integrands, we can replace $dk$ by $\Delta k$, $f$ by its mean value $\overline{f}$ and $k,k'$ by $k_0$. Equation \eqref{eq:norm_integral_regularized} then becomes
\begin{eqnarray}
\int d^3r S_{k_0}(\vec{r})\vec{E}_{nlm}(\vec{r})\vec{E}_{n'l'm'}(\vec{r}) &\approx& \Delta k \cdot \overline{f}^2 \cdot \frac{\pi\epsilon_b}{k_0^2} \left(Z_{\text{ii}}^{(n,l,m,n',l',m')}+Z_{\text{oo}}^{(k_0,n,l,m,k_0,n',l',m')}\right) + \nonumber \\
& & \left(\Delta k\right)^2 \cdot \overline{f}^2 \cdot \overline{F}^{(k_0,n,l,m,k_0,n',l',m')}(R) \quad . \label{eq:norm_regularized}
\end{eqnarray}
In taking the limit $\Delta k \rightarrow 0$ we now have to take care that Eq.~\eqref{eq:norm_regularized}, which describes the energy of the electromagnetic field, takes on a finite value. Since $Z_{\text{ii}}$, $Z_{\text{oo}}$, and $F$ are all finite, we see that $\Delta k \rightarrow 0$ has to be taken in such a way that $\Delta k \cdot \overline{f}^2$ remains finite. This leads to the term containing $F$ approaching zero linear in $\Delta k$. We therefore find that, in a physically meaningful limit, only these terms that contain a suitable $\delta$-distribution in the normalization integral contribute.

The normalization integrals for the electromagnetic field in the presence of a localized scatterer embedded in a lossless, non-dispersive background medium hence read
\begin{equation} \label{eq:main_result_spherical_harmonics}
W(k,n,l,m,k',n',l',m') = \frac{\pi\epsilon_b}{k^2}\left(Z_{\text{ii}}^{(n,l,m,n',l',m')}+Z_{\text{oo}}^{(k,n,l,m,k',n',l',m')}\right)\delta(k-k') \quad .
\end{equation}
At this point a few important features of Eq.~\eqref{eq:main_result_spherical_harmonics} are worth pointing out. First of all, only the surface integrals $Z_{\text{ii}}$ and $Z_{\text{oo}}$ contribute to the normalization integrals and therefore only information about the far field radiation pattern is required. This is in accordance to Ref.~\cite{Devaney1974}, where it was shown that the field distribution in the entire space excluding the scattering region can be inferred from the far field radiation pattern only. Secondly, no cross terms between input and output fields occur anymore in Eq.~\eqref{eq:main_result_spherical_harmonics}. This relates nicely to the simple physical picture that input and output are temporally separated and therefore can be mathematically treated separatly. Thirdly, the normalization integral \eqref{eq:main_result_spherical_harmonics} is independent of the imaginary sphere radius $R$. This allows us to take the limit $R\rightarrow\infty$ and hence restore all orders in $l$, as was announced in the paragraph following Eq.~\ref{eq:norm_integral}. The normalization integrals $Z$ can be evaluated in terms of the multipole coefficients
\begin{eqnarray}
Z_{\text{ii}}^{(n,l,m,n',l',m')} &=& \delta_{nn'}\delta_{ll'}\delta_{mm'}\left|b_{nlm}\right|^2 \quad , \\
Z_{\text{oo}}^{(k,n,l,m,k',n',l',m')} &=& \sum_{n''l''m''}a_{n''l''m''}^{(k,n,l,m)\dagger}a_{n''l''m''}^{(k',n',l',m')} \quad . \label{eq:norm_multipole_representation}
\end{eqnarray}

\subsection{Expansion in plane waves}
\label{sec:plane_waves}

We will now proceed to treat the important case of plane wave illumination.

Assuming that our system is illuminated by a single plane wave with wavevector $\vec{k}$ and polarization $\sigma$, we write the resulting field in the entire space as
\begin{equation} \label{eq:total_field}
\vec{E}_{\vec{k},\sigma}(\vec{r}) = \vec{E}_{\vec{k},\sigma,0} \ee^{\ii\vec{k}\cdot\vec{r}} + \vec{E}^{(\vec{k},\sigma)}_{S}(\vec{r}) \quad  ,
\end{equation}
where the scattered field $\vec{E}^{(\vec{k},\sigma)}_{S}(\vec{r})$ is defined as having no incoming components. Since the discussion in Section \ref{sec:spherical_harmonics} entails only the far field contributing to the normalization integral, we can use the asymptotic form of the scattered field
\begin{equation} \label{eq:scattered_field}
\vec{E}^{(\vec{k},\sigma)}_{S}(\vec{r}) \approx \vec{A}^{(\vec{k},\sigma)}(\hat{r})\frac{\ee^{\ii kr}}{kr} \quad .
\end{equation}
After inserting Eqs.~\eqref{eq:total_field} and \eqref{eq:scattered_field} into \eqref{eq:normalisation_integral}, we proceed by splitting the integration region into outside and inside contributions as before. Since the inside parts were shown to yield no contribution in Section \ref{sec:spherical_harmonics}, we immediately drop them to arrive at
\begin{align}\label{eq:norm_integral_plane_wave}
\frac{1}{\epsilon_b}W(\vec{k}, \sigma, \vec{k'}, \sigma') &= W_1(\vec{k}, \sigma, \vec{k'}, \sigma') + W_2(\vec{k}, \sigma, \vec{k'}, \sigma') + W_3(\vec{k}, \sigma, \vec{k'}, \sigma') + W^*_3(\vec{k'}, \sigma', \vec{k}, \sigma) \quad , \\
W_1(\vec{k}, \sigma, \vec{k'}, \sigma') &= \int d^3r \vec{E}^{\dagger}_{\vec{k},\sigma,0} \cdot \vec{E}_{\vec{k'},\sigma',0} \ee^{-\ii(\vec{k}-\vec{k'})\cdot\vec{r}} \quad , \nonumber \\
W_2(\vec{k}, \sigma, \vec{k'}, \sigma') &= \int d^3r \vec{A}^{(\vec{k},\sigma)\dagger}(\hat{r}) \cdot \vec{A}^{(\vec{k'},\sigma')}(\hat{r}) \frac{\ee^{-\ii(k-k')r}}{(kr)^2} \quad , \nonumber  \\
W_3(\vec{k}, \sigma, \vec{k'}, \sigma') &= \int d^3r \vec{E}^{\dagger}_{\vec{k},\sigma,0} \cdot \vec{A}^{(\vec{k'},\sigma')}(\hat{r}) \frac{\ee^{-\ii(k\cos\theta-k')r}}{kr} \quad , \nonumber
\end{align}
where $\theta$ denotes the angle between $\vec{k}$ and $\vec{r}$. The first integration above can be easily performed
\begin{equation}
W_1(\vec{k}, \sigma, \vec{k'}, \sigma') = \left|\vec{E}_{\vec{k},\sigma,0}\right|^2 (2\pi)^3 \delta^{(3)}(\vec{k}-\vec{k'}) \quad .
\end{equation}
For the remaining terms the radial integral is the only possible source of non-finite contributions. We will, therefore, only concern ourselves with integrations over $r$. The second integral in Eq.~\eqref{eq:norm_integral_plane_wave} immediately yields
\begin{equation}
W_2(\vec{k}, \sigma, \vec{k'}, \sigma') = \frac{1}{k^2} \left(\pi\delta(k-k')-\text{P}\frac{\ii}{k-k'}\right)\int d\Omega \vec{A}^{(\vec{k},\sigma)\dagger}(\hat{r}) \cdot \vec{A}^{(\vec{k'},\sigma')}(\hat{r}) \quad ,
\end{equation}
while the last two terms need some more care. Partial integration over $\cos \theta$ leads to
\begin{align}
W_3&(\vec{k}, \sigma, \vec{k'}, \sigma') \nonumber \\ 
=& \int_0^\infty dr \int d\phi \left[\vec{E}^{\dagger}_{\vec{k},\sigma,0} \cdot \vec{A}^{(\vec{k'},\sigma')}(\hat{r})\frac{\ii}{k^2}\ee^{-\ii(k\cos\theta-k')r}\right]_{\cos\theta=-1}^1 - \nonumber \\
& \int_0^\infty dr \int d\Omega \vec{E}^{\dagger}_{\vec{k},\sigma,0} \cdot \frac{\partial}{\partial \cos\theta}\left(\vec{A}^{(\vec{k'},\sigma')}(\hat{r})\right)\frac{\ii}{k^2}\ee^{-\ii(k\cos\theta-k')r} \nonumber \\
=& \frac{2\pi\ii}{k^2} \vec{E}^{\dagger}_{\vec{k},\sigma,0} \cdot \vec{A}^{(\vec{k'},\sigma')}(\hat{k}) \left[\pi\delta(k-k')-\text{P}\frac{\ii}{k-k'}\right] - \nonumber \\
& \frac{2\pi\ii}{k^2} \vec{E}^{\dagger}_{\vec{k},\sigma,0} \cdot \vec{A}^{(\vec{k'},\sigma')}(-\hat{k}) \left[\pi\delta(k+k')+\text{P}\frac{\ii}{k+k'}\right] - \nonumber \\
& \frac{\ii}{k^2} \int d\Omega \vec{E}^{\dagger}_{\vec{k},\sigma,0} \cdot \frac{\partial}{\partial \cos\theta}\left(\vec{A}^{(\vec{k'},\sigma')}(\hat{r})\right) \left[\pi\delta(k\cos\theta-k')-\text{P}\frac{\ii}{k\cos\theta-k'}\right] \quad . \label{eq:plane_wave_normalisation_integral}
\end{align}
Before we proceed with the physical argument used in Sec.~\ref{sec:spherical_harmonics}, we need to examine the last term in \eqref{eq:plane_wave_normalisation_integral} more closely. We see that for $k=k'$ the integrand can only yield infinite contributions for $\cos\theta = 1$. But since $\{1\}$ is a subset of $[-1, 1]$ of measure $0$, the contribution from the integral can at most be finite and we can therefore discard the term, just as we did with other finite terms.

We now define a physical field as a superposition of plane waves
\begin{equation}
\vec{E}(\vec{r}) = \int d^3k f(\vec{k}) E_{\vec{k},\sigma}(\vec{r}) \quad , \label{eq:physical_plane_wave}
\end{equation}
where $f$ only has support on $I \times \Omega_{\vec{k}}$ with $I = [k_0-\Delta k/2, k_0+\Delta k/2]$ and $\Omega_{\vec{k}}$ a solid angle of size $\Delta\Omega_{\vec{k}}$. By inserting \eqref{eq:physical_plane_wave} into \eqref{eq:norm_integral_plane_wave} and, in analogy to Sec.~\ref{sec:spherical_harmonics}, making the substitutions $k,k' \rightarrow k_0$, $dk \rightarrow \Delta k$, $d\Omega_{\vec{k}} \rightarrow \Delta \Omega_{\vec{k}}$ and $f \rightarrow \overline{f}$, where $\overline{f}$ is the mean value of $f$, we find $W_1 \propto \Delta k \cdot \Delta \Omega_{\vec{k}} \cdot \overline{f}^2$, $W_2 \propto \Delta k \cdot \left(\Delta \Omega_{\vec{k}}\right)^2 \cdot \overline{f}^2$ and $W_3 \propto \Delta k \cdot \left(\Delta \Omega_{\vec{k}}\right)^2 \cdot \overline{f}^2$. Letting $\Delta k \rightarrow 0$ and $\Delta\Omega_{\vec{k}} \rightarrow 0$ while enforcing finiteness of the result, just as in Sec.~\ref{sec:spherical_harmonics}, we find that only $W_1$ stays finite while $W_2$ and $W_3$ tend to zero linearly in $\Delta \Omega_{\vec{k}}$.

We therefore find for the normalization integral in the case of plane wave illumination
\begin{equation}
W(\vec{k},\sigma,\vec{k'},\sigma') = (2\pi)^2\epsilon_b\vec{E}^{\dagger}_{\vec{k},\sigma,0}\cdot\vec{E}_{\vec{k'},\sigma',0} \delta^{(3)}(\vec{k}-\vec{k'}) \quad .
\end{equation}
This equation tells us that the solutions corresponding to incident plane waves with $\vec{k}\neq\vec{k'}$ are already orthogonal to each other. When $\vec{k}=\vec{k'}$, the solutions are orthogonal if the polarization vectors of the incident plane waves are orthogonal. Furthermore, the correctly normalized fields are of the form
\begin{equation} \label{eq:plane_wave_result}
\vec{E}_{\vec{k},\sigma}(\vec{r}) = \hat{e}_{\vec{k},\sigma} \sqrt{\frac{\hbar\omega}{(2\pi)^32\epsilon_0\epsilon_b}} \ee^{\ii\vec{k}\cdot\vec{r}} + \vec{E}^{(\vec{k},\sigma)}_{S}(\vec{r}) \quad ,
\end{equation}
where the polarization vectors $\hat{e}_{\vec{k},\sigma}$ are normalized. This means that in order to achieve correct field orthonormalization in any case where a plane wave is incident on a localized scatterer one simply has to fix the amplitude of the plane wave according to \eqref{eq:plane_wave_result}.

The result \eqref{eq:plane_wave_result} might be surprising at first, since the normalization does not depend on the scattered field. This is due to the nature of plane waves. In Sec.~\ref{sec:spherical_harmonics} we showed that the contribution from within the scattering region can be ignored compared to the contribution from the infinitely extended multipole field, while in Sec.~\ref{sec:plane_waves} we showed that the contribution of the outgoing multipoles decaying as $r^{-1}$ can be ignored compared to the contribution of the plane wave extending over whole space. The normalization of plane waves is therefore insensitive against localized changes in permittivity due to their unlocalized nature.

\section{Numerical examples}
\label{sec:numerics}

To demonstrate the relevance and broad applicability of the proposed normalization scheme, we include here a series of numerical calculations that require correct normalization of the fields to contain the energy of a single photon. Please note that this list of applications is by no means exhaustive, but representative for the current research in quantum optics.

\subsection{Single-photon field amplitude and vacuum fluctuations}

\begin{figure}
\begin{center}
\includegraphics[width=0.5\textwidth]{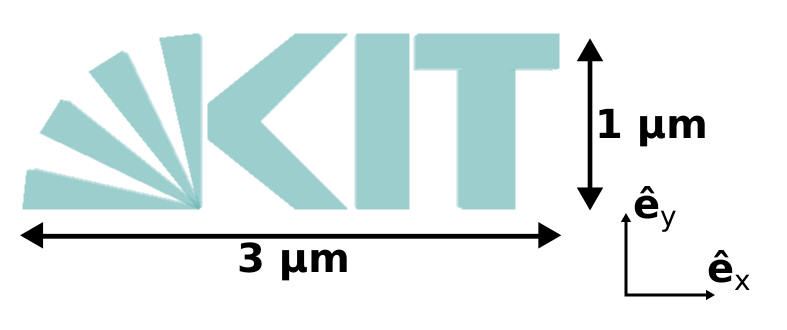}
\includegraphics[width=0.8\textwidth]{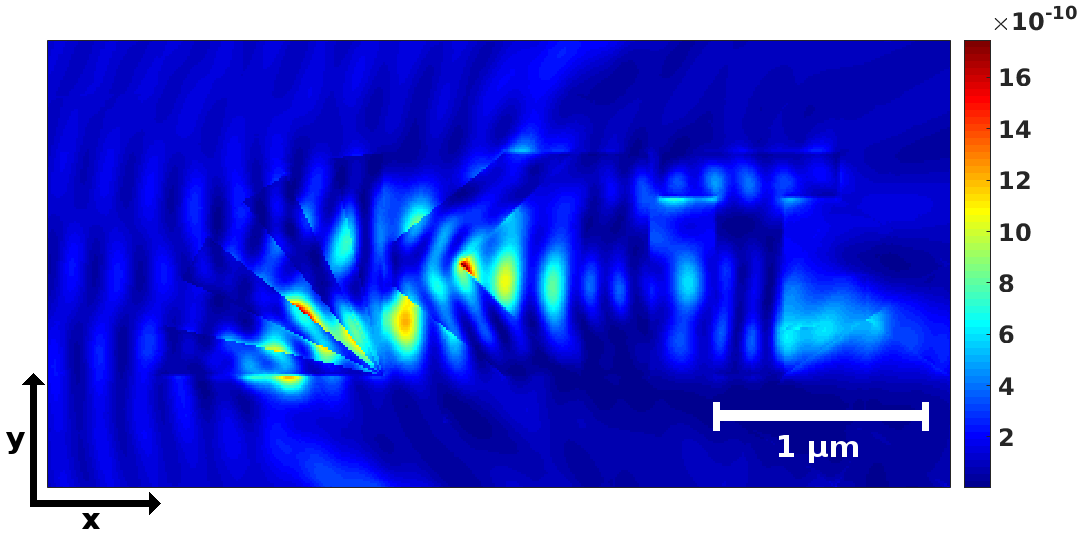}
\end{center}
\caption{Top: The scatterer under consideration when viewed along the $z$-direction in which the scatterer has a finite extent from $z = 0$ to  $1~\mu\text{m}$. Bottom: Distribution of the quantity $I_{\vec{k},\sigma}(\vec{r})$ in a plane of $z=0.5~\mu\text{m}$. Here we have chosen $\vec{k} = 1.668\cdot10^7~\text{m}^{-1}\cdot\hat{e}_x$ and polarization along $\hat{e}_y$. The scatterer is made from a dielectric material with a relative permittivity $\epsilon = 2.25$ and is embedded in vacuum. $I_{\vec{k},\sigma}(\vec{r})$ is given in SI units. \label{fig:kit_scattering}}
\end{figure}
As an elementary first example we are interested in the quantity
\begin{equation}
I_{\vec{k},\sigma}(\vec{r}) = \bra{1_{\vec{k},\sigma}}\vec{E}^{(-)}(\vec{r})\vec{E}^{(+)}(\vec{r})\ket{1_{\vec{k},\sigma}} \quad , \label{eq:single_photon_amplitude}
\end{equation}
where $\vec{E}^{(\pm)}(\vec{r})$ denotes the positive and negative frequency parts of the electric field operator \cite{Vogel1994} and $\ket{1_{\vec{k},\sigma}}=a^\dagger_{\vec{k},\sigma}\ket{0}$ is a single-photon state in the plane wave basis. $I_{\vec{k},\sigma}(\vec{r})$ is of fundamental importance in detector theory \cite{Vogel1994} and allows us to calculate single-photon detection probabilities, if a suitable detector model is known. Evaluation of the left hand side of Eq.~\eqref{eq:single_photon_amplitude} shows that $I_{\vec{k},\sigma}(\vec{r})$ is just the square norm of the normalized electric field strength for an incident plane wave of wavevector $\vec{k}$ and polarization $\sigma$.

Figure \ref{fig:kit_scattering} shows the single-photon intensity $I_{\vec{k},\sigma}(\vec{r})$ in the vicinity of a complex scatterer for particular values of $\vec{k}$ and $\sigma$. The scatterer corresponds to the logo of the institute of the authors to demonstrate that indeed an arbitrary object can be considered. The numerical calculations have been performed with the commercially available FEM software JCMsuite. To our knowledge this is the first time that $I_{\vec{k},\sigma}(\vec{r})$ was calculated without using an analytical solution to Maxwell's equations.

It is worth noting that
\begin{equation}
I_{\vec{k},\sigma}(\vec{r}) = \bra{0}\vec{E_{\vec{k},\sigma}}(\vec{r})\vec{E_{\vec{k},\sigma}}(\vec{r})\ket{0} = \bra{0}
[\vec{E^{(+)}_{\vec{k},\sigma}}(\vec{r})+\vec{E^{(-)}_{\vec{k},\sigma}}(\vec{r})][\vec{E^{(+)}_{\vec{k},\sigma}}(\vec{r})+\vec{E^{(-)}_{\vec{k},\sigma}}(\vec{r})]\ket{0} \quad ,
\end{equation}
i.e. $I_{\vec{k},\sigma}(\vec{r})$ also describes the vacuum  fluctuations of the electromagnetic field. Since vacuum fluctuations give rise to Casimir forces between objects \cite{Novotny2012}, our normalization scheme paves the way for the calculation of Casimir forces in arbitrary optical systems.

\subsection{Coupling to a two-level system}
\label{subsec:coupling}

A common problem in cavity quantum electrodynamics is the calculation of the coupling strength $\kappa$ between a single two-level quantum system at a point in space $\vec{r}_{\text{TLS}}$ and 
a resonant cavity mode of a scatterer, excited either by an external illumination, e.g. a plane wave, or by the two-level quantum system itself. Here we define a cavity mode according to the discussion in \cite{Placeholder}, where the coupling constant is calculated as
\begin{equation}
\kappa = \sqrt{\frac{C}{4\pi}}\frac{\pi}{c_0^{3/2}}\omega_0\sqrt{2\Gamma} \frac{\vec{E}_{\vec{k_0}}(\vec{r_{\text{TLS}}})\cdot\vec{d}}{\hbar} \quad , \label{eq:kappa_ff}
\end{equation}
where $\omega_0$ is the mode frequency, $\Gamma$ the full width at half maximum, $\vec{d}$ the dipole moment of the two level system, $\vec{E}_{\vec{k_0}}$ the field strength at resonance and
\begin{equation}
C = \int_{|\vec{k}| = \omega_0/c_0} d\Omega_k \frac{|\vec{E}_{\vec{k}}(\vec{r}_{\text{TLS}})|^2}{|\vec{E}_{\vec{k_0}}(\vec{r}_{\text{TLS}})|^2} \quad ,
\end{equation}
the normalized integrated response per incidence angle, where $\vec{E}_{\vec{k}}$ denotes the field for plane wave illumination with wavevector $\vec{k}$. Please note that we have not attached a polarization index to the electric field, since we assume that only one choice of polarization leads to a resonance, as is the case for the geometry discussed below.

As a point of reference for the scheme presented here we used a technique commonly encountered in the literature, where a finite quantization volume $V$ is chosen and the field is normalized according to \cite{Vernooy1997, Buck2003, Slowik2013, Esteban2014}
\begin{equation} \label{eq:numerical_integral}
\frac{1}{4}\int_{V} d^{3}r\frac{\text{d}}{\text{d}\omega}\left[\omega\text{Re}\left\{\epsilon_0\epsilon(\vec{r},\omega)\right\}\right]\vec{E}^{\dagger}(\vec{r})\cdot\vec{E}(\vec{r}) = \frac{\hbar\omega}{2} \quad .
\end{equation}

Here, we choose the computational domain of the numerical simulations as the quantization volume. The coupling constant is then calculated according to \cite{Vogel1994}
\begin{equation}
\kappa' = \frac{\vec{E}(\vec{r_{\text{TLS}}})\cdot\vec{d}}{\hbar} \quad . \label{eq:kappa_cd}
\end{equation}
Please note the resemblance between Equations \eqref{eq:kappa_ff} and \eqref{eq:kappa_cd}, which only differ by a factor of dimension $\text{m}^{-3/2}$. This factor can be identified with the square root of the mode volume in $\vec{k}$-space.

\begin{figure}
\begin{center}
\includegraphics[width=.49\textwidth]{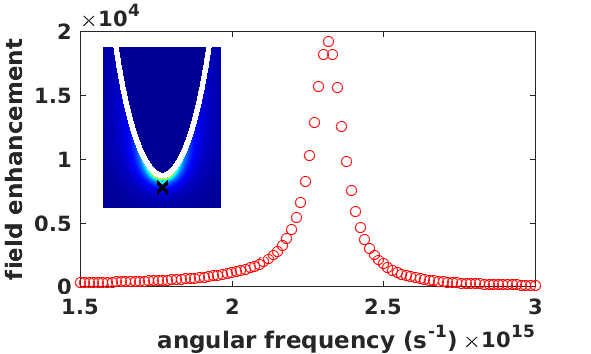}
\includegraphics[width=.49\textwidth]{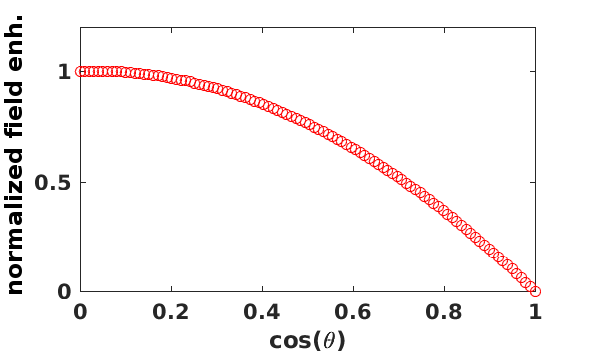}
\end{center}
\caption{Frequency (left) and incidence angular (right) spectrum at resonance of a gold prolate ellipsoid with major radius $r_a = 60~\text{nm}$ and minor radius $r_b = 10~\text{nm}$. Plotted here is the square of the amplitude enhancement (relative to the incident plane wave's amplitude) at a distance of $1~\text{nm}$ from the tip (black cross). The inset shows the field intensity distribution at resonance, the white line marking the surface of the rod.}
\label{fig:plasmonic_rod}
\end{figure}

We will now proceed to calculate the coupling strength $\kappa$ between a gold nanoantenna and a two-level system placed $1~\text{nm}$ away from the tip of the antenna with the two approaches. The nanoantenna has the shape of a prolate ellipsoid with major radius $r_a = 60~\text{nm}$ and minor radius $r_b = 10~\text{nm}$. The transition dipole moment of the two-level system is aligned along the antenna axis and we assume $d/\hbar = 10^5~\text{C}\text{N}^{-1}\text{s}^{-1}$, which is close to realistic values \cite{Eliseev2000}. Systems similar to the one described here have been considered for various applications \cite{Maksymov2011, Slowik2013}.

Figure \ref{fig:plasmonic_rod} shows the field enhancement relative to the amplitude of the incident plane wave at the position of the two-level system. The simulations have been performed with JCMsuite. The computational domain is a cylinder with a radius and height of $300~\text{nm}$. The coupling constant is calculated to be $\kappa\approx4.81\cdot10^{12}~\text{s}^{-1}$ when using the proposed normalization scheme together with Eq.~\eqref{eq:kappa_ff} and $\kappa'\approx1.85\cdot10^{12}~\text{s}^{-1}$ when using Eqs.~\eqref{eq:numerical_integral} and \eqref{eq:kappa_cd}, showing a deviation of roughly a factor of $3$. Since plasmonic nanoantennas are capable of just reaching the strong coupling regime of light-matter interaction, a factor of $3$ can mean the difference between weak and strong coupling. The predictions made by using the two normalization schemes might therefore not only differ quantitatively, but also qualitatively.

The deviation is due to two effects, shifting the field strength in opposite directions when using the normalization integral \eqref{eq:numerical_integral}. On the one hand the high field enhancement increases the value of the normalization integral \eqref{eq:numerical_integral}, but on the other hand the small mode volume decreases it. In the end these effects cancel each other well enough to at least yield a value of the right order of magnitude, but this is by no means obvious. For more sophisticated plasmonic systems one can therefore expect a similar or even greater discrepancy between the two schemes.

\subsection{Spontanous emission rate}

As a final example we discuss the modified spontaneous emission rate of an emitter placed at a distance of $5~\text{nm}$ from the tip of the plasmonic antenna discussed in Section \ref{subsec:coupling}. To this end, we use the formula \cite{Sauvan2013}
\begin{equation}
\Gamma(k) = \frac{2}{\hbar}\Im\left[\vec{p}^*(k)\cdot\vec{E}_k(\vec{r}_{\text{em}})\right] = \frac{2}{\hbar}|\vec{E}_k(\vec{r}_{\text{em}})|^2\Im[\alpha^*(k)] \quad , \label{eq:emission_rate}
\end{equation}
where $\Gamma(k) \cdot dk$ is the emission rate into the wavevector interval $[k, k+dk]$, $\vec{E}_k(\vec{r}_{\text{em}})$ is the normalized total electric field of an electric point dipole of frequency $\omega$ at the position $\vec{r}_{\text{em}}$ of the dipole, $\vec{p}(k)=\alpha(k)\vec{E}_k(\vec{r}_{\text{em}})$ is the dipole moment as described by the frequency-dependent polarizability $\alpha(\omega)$ and $\Im$ denotes the imaginary part. To obtain the total emission rate of the emitter one has to integrate Eq.~\eqref{eq:emission_rate} over all frequencies. Assuming that the polarizability is of Lorentzian shape and much narrower than the antenna mode
\begin{equation}
\alpha(k) = \alpha(k_0)\frac{(\gamma/2)^2}{(k-k_0)^2+(\gamma/2)^2} \quad ,
\end{equation}
 we find for the total emission rate
\begin{equation}
\Gamma = \int_0^\infty dk\Gamma(k) = \frac{\pi}{\hbar}\gamma|\vec{E}_{k_0}(\vec{r}_{\text{em}})|^2\Im[\alpha^*(k_0)] \quad .
\end{equation}
Please note that the real part of the electric field strength diverges at the position of the emitter, while the imaginary part stays finite. However, since the divergent part is independent of the scattering geometry, it can be seen as a vacuum contribution. This is similar to the finite but constant vacuum term in the Hamiltonian that arises during quantization of the electromagnetic field \cite{Vogel1994}. Such infinite terms can be absorbed into renormalization constants that lead from bare to dressed (or physical) quantities \cite{Ryder1996}. Since $\alpha(k)$ is already the physically observable value of the polarizability, the divergent terms don't need to be considered in the calculations.

The numerical simulations were performed with COMSOL Multiphysics and the normalization of the electric field was achieved by multipole expansion as described in Section \ref{sec:spherical_harmonics}. To obtain a reasonable value of $\Im[\alpha^*]$ we use the results in \cite{Jentschura2015}, where it is shown that $\Im[\alpha^*]/\alpha \approx \alpha_{\text{QED}}^3$ with the fine-structure constant $\alpha_{\text{QED}}\approx\frac{1}{137}$. Together with the polarizabilities calculated in \cite{Goldman1989} this leads us to an estimated value of $\Im[\alpha^*(k_0)]/\hbar = 10^{-13}~\text{C}^2\text{m}\text{N}^{-1}$. For the spectral width of the polarizability we assumed a value of $\gamma = 10^4~\text{m}^{-1}$.

\begin{figure}
\begin{center}
\includegraphics[width=.6\textwidth]{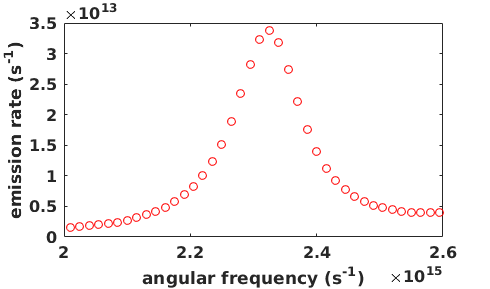}
\end{center}
\caption{Emission rate of an emitter placed $5~\text{nm}$ from the tip of a plasmonic nanoantenna as a function of the emission frequency. Maximum emission occurs at the resonance frequency of the nanoantenna.}
\label{fig:emission_rate}
\end{figure}

The results are shown in Fig. \ref{fig:emission_rate}, where the modified spontanous emission rate of an emitter was calculated as a function of its emission frequency. Maximum emission is achieved at the resonance frequency of the nanoantenna, as expected.

\section{Conclusion}
\label{sec:conclusion}

We have proposed a novel scheme for the normalization of electromagnetic scattering modes for arbitrary localized scatterers embedded in a uniform, lossless background medium. To this end, we have used analytical far field solutions of Maxwell's equations and a physical argument concerning the physical meaning of infinitely extended waves. We have given easy-to-implement methods and formulas for the normalization and orthogonalization of arbitrary solutions to scattering problems. We have also treated the important case of plane wave scattering solutions and solutions that consider a given mutipolar field as illumination. Our scheme was demonstrated in several examples and important quantities such as light-matter coupling strength and spontaneous emission rate where retrieved for realistic setups.

Correct field normalization is the basis of all quantum optical calculations and our approach allows to treat it in a numerically feasible and rigorous way. This is especially useful when considering plasmonic systems, which commonly feature strong field enhancement and a small mode volume. Recent experimental findings \cite{Marinica2015, Chikkaraddy2016} have reignited interest in plasmonic systems reaching the strong coupling regime \cite{Hummer2013}, promising new quantum technological applications at room temperature \cite{Jacob2012, Tame2013} due to the preservation of quantum coherence \cite{Torma2015}.

\section{Acknowledgements}
The authors acknowledge support from the DFG project RO 3640/4-1. J.S. acknowledges support from the Karlsruhe School of Optics and Photonics (KSOP). The authors also acknowledge support from the company JCMWave for providing us the software used in some of the simulations.
\end{document}